\documentclass[%
reprint,
amsmath,amssymb,
aps,
nofootinbib,
pre,
twocolumn
]{revtex4-1}

\usepackage{mathbbol}
\usepackage{graphicx}
\usepackage{dcolumn}
\usepackage{bm}

\newcommand{\CO}{{\cal O}}

\def\IR{{\mathbb R}}

\def\rB{{\rm{B}}}
\def\rF{{\rm{F}}}

\newcommand{\mD}{\mathsf{D}}

\newcommand{\re}{{\rm e}}
\newcommand{\ri}{{\rm i}}
\newcommand{\rd}{{\rm d}}

\newcommand{\be}{\begin{equation}}
\newcommand{\ee}{\end{equation}}
\newcommand{\ba}{\begin{aligned}}
\newcommand{\ea}{\end{aligned}}
\newcommand{\ben}{\begin{eqnarray}\displaystyle}
\newcommand{\een}{\end{eqnarray}}

\newcommand{\figref}[1]{Fig.~\protect\ref{#1}}

\begin{document}

\title{Exact perturbative results for the Lieb--Liniger and Gaudin--Yang models}

\author{Marcos Mari\~no and Tom\'as Reis}
\affiliation{D\'epartement de Physique Th\'eorique et Section de Math\'ematiques, Universit\'e de Gen\`eve, CH 1211, Geneva, Switzerland}

\begin{abstract}
We present a systematic procedure to extract the perturbative series for the ground state energy density in the 
Lieb--Liniger and Gaudin--Yang models, starting from the Bethe ansatz solution. This makes it possible 
to calculate explicitly the coefficients of these series and to study their large order behavior. We find that both series diverge 
factorially and are not Borel summable. In the case of the 
Gaudin--Yang model, the first Borel singularity is determined by the non-perturbative energy gap. This provides a new perspective 
on the Cooper instability.

\end{abstract}

\pacs{Valid PACS appear here}
\maketitle

\section{\label{intro}Introduction}

The Lieb--Liniger (LL) and Gaudin--Yang (GY) models describe a gas of one-dimensional bosons (respectively, spin $1/2$ fermions) with a delta function interaction. 
As shown in the pioneering papers \cite{ll,gaudin,yang}, these models are exactly solvable with the Bethe ansatz, and for this reason they are 
excellent laboratories to test our understanding of interacting quantum gases. In addition, they can be realized in the laboratory, leading to a fascinating 
interplay between integrable quantum models and experiment (see \cite{guan-boson,guan-review,bthierry,bdz} for reviews and references). 

 Typically, exact solutions can be used to generate series in the coupling constant of the model, providing in this way a 
 powerful test of perturbative methods. However, as already noticed in \cite{ll}, the weakly coupled regime of the LL and GY models turns out to be 
very difficult to analyze by using the Bethe ansatz equations. Although the exact solution of these models has been known for fifty years, 
and in spite of many efforts \cite{popov, iw-gy,tw-ll,tw1,tw2}, only the first three terms of the perturbative series for the 
ground state energy density of these models have been derived. More recently, precise numerical calculations have made it possible to guess the analytic form of some additional coefficients 
\cite{prolhac, lang, zr-numerical}. 

In this paper we solve this longstanding problem and give a systematic procedure to derive the perturbative expansion for the ground 
state energy density of these models, building on previous results in \cite{volin, volin-thesis}. 
The procedure is recursive and can be easily implemented in a symbolic program. In this way we are able to compute analytically the 
perturbative series up to order $50$ in the coupling constant. We  
use our results to determine their large order behavior, which is an important source of non-perturbative information. 
We find that, in both cases, the series are factorially divergent. In the LL and attractive GY models, the series are non-alternating and therefore non-Borel summable. 
In the case of the attractive Gaudin--Yang model, whose ground state is a BCS-like state, 
we show that the associated singularity in the Borel plane is closely related to the 
superconducting energy gap.

\section{\label{ll-gy}The Lieb--Liniger and Gaudin--Yang models}

The Hamiltonian for the LL and GY models is given by (we set $\hbar=2m=1$)
\be
\label{h-llgy}
H_N=-\sum_{j=1}^N {\partial^2 \over \partial x_j^2}+ 2 c \sum_{ 1\le i<j\le N} \delta(x_i-x_j).  
\ee
In the LL case we have a repulsive interaction with $c>0$, while in the GY case we will consider the attractive case $c<0$. 
We will focus on the ground state energy density of the models: $E= \lim_{L \rightarrow \infty} E_0(N, L)/L$, where $E_0(N,L)$ is the ground state energy for $N$ particles 
on the interval $[0,L]$ with periodic boundary conditions. We consider the thermodynamic limit in which $L , N\rightarrow \infty$ while the density $n=N/L$ remains fixed. The value of $E$ 
can be obtained from the solution of a linear integral equation for the density of Bethe roots, which is supported on an interval $[-B,B]$ (see \cite{sb-book} for an 
excellent presentation of the Bethe ansatz solution of these models). In the LL model, the integral equation reads
\be
\label{ll-ie}
{f_\rB(x) \over 2} -{1\over 2 \pi} \int_{-B}^B {f_\rB(y)  \rd y \over (x-y)^2+1} =1. 
\ee
The dimensionless coupling constant $\gamma=c/n$ is given by 
\be
\label{b-gam}
{1\over \gamma}= {1\over 4 \pi} \int_{-B}^B f_\rB(x) \rd x, 
\ee
while 
\be
\label{b-e}
e_B(\gamma)={E \over n^3}= {\gamma^3 \over  4 \pi} \int_{-B}^B x^2 f_\rB(x) \rd x. 
\ee
In the GY model, the integral equation reads
\be
\label{gy-ie}
{f_\rF(x) \over 2} +{1\over 2 \pi} \int_{-B}^B {f_\rF(y) \rd y \over (x-y)^2+1} =1.  
\ee
The dimensionless coupling constant $\gamma=|c|/n$ is now given by 
\be
\label{f-gam}
{1\over \gamma}= {1\over \pi} \int_{-B}^B f_\rF(x) \rd x, 
\ee
while 
\be
\label{f-e}
e_\rF(\gamma)={E \over n^3}= -{\gamma^2 \over 4}+ {\gamma^3 \over   \pi} \int_{-B}^B x^2 f_\rF(x) \rd x. 
\ee
Note that the ground state energy density depends on the coupling constant through $B$, the endpoint of the support of the 
distribution. The strong coupling limit of these equations, $\gamma \rightarrow \infty$, can be studied systematically in a rather straightforward way (see 
e.g. \cite{guan-review,zr}). However, the weak coupling limit $\gamma \rightarrow 0$ corresponds to $B \rightarrow \infty$, which is a singular limit of the 
integral equations. For this reason, extracting systematically the perturbative expansion from these integral equations has been an open problem 
since Lieb and Liniger introduced their equation in 1963. The known analytic results are \cite{tw-ll,tw1,tw2}
\be
\ba
e_\rB(\gamma)&= \gamma -{4 \gamma^{3/2} \over 3\pi}+\left( {1\over 6}- {1\over \pi^2} \right) \gamma^2+ \CO\left(\gamma^{5/2} \right), \\
e_\rF(\gamma)&= \frac{\pi ^2}{12}-\frac{\gamma}{2}-\frac{\gamma^2}{12}+\CO(\gamma^3). 
\ea
\ee

\section{\label{exact}Exact perturbative results}

The method we will use to determine these series goes back to the work of Hutson on the related mathematical 
problem of the circular 
plate condenser \cite{hutson}, and to the work of 
Popov on the LL equation \cite{popov}. It was substantially improved by D. Volin in \cite{volin,volin-thesis}, where he applied it to 
integrable field theories in two dimensions (see also \cite{ksv}). We will explain now this method in some detail, since it is likely to have other applications. 
It has three ingredients. First, one rewrites the integral equation as a difference equation for the 
{\it resolvent} of the density of Bethe roots. Second, one studies two different regimes for the equation: the bulk regime near the origin, and the edge regime near the 
endpoint of the support of the distribution. One then matches asymptotic expansions in the two regimes (this was already the main idea in \cite{hutson,popov}). The third ingredient 
involves finding an explicit solution for the difference equation in the edge regime. 

Let us then first introduce the resolvent of the densities:
 \be
 R(x)= \int_{-B}^B {f(x') \over x-x'} \rd x'. 
 \ee
 (When no subscript is introduced, it is assumed that the equations are valid for both LL and GY models.) 
 The resolvent is analytic in the complex $x$-plane, except for a discontinuity in the interval $[-B, B]$, given by 
 \be
 \label{disc}
 f(x)=-{1\over 2 \pi \ri} \left( R^+(x)-R^-(x) \right), 
 \ee
 where $R^\pm (x)=R(x\pm \ri \epsilon)$. Note that $R(x)$ is an odd function on the real axis away from $[-B, B]$ 
 (its discontinuity (\ref{disc}) is an even function of $x$). From its definition we deduce that 
 \be
 \label{asym}
 R(x)= \sum_{k \ge 0} \langle x^k\rangle  x^{-k-1} , \qquad\langle x^k\rangle= \int_{-B}^B f(x) x^k \rd x. 
 \ee
In particular, we can compute the coupling constant $\gamma$ as a function of $B$ from the residue at infinity of the resolvent. By writing 
\be
{1\over x^2+1}= -{1\over 2 \ri}\left( \mD - \mD^{-1}\right){1\over x},  \qquad \mD= \re^{\ri \partial_x}, 
\ee
 is now easy to see that the LL and Gaudin's integral equation can be transformed into difference equations for the corresponding resolvents:
\be
\label{gy-ll}
\ba
(1-\mD) \tilde R^+_\rB (x)- (1- \mD^{-1}) \tilde R^-_\rB (x) &=0, \\
(1+\mD) R^+_\rF (x)- (1+ \mD^{-1}) R^-_\rF (x) &=-4 \pi \ri, \ea
\ee
where $\tilde R_\rB(x)= R_\rB(x) -2 \pi x$. The subtraction of the linear term takes into account the known leading order behavior 
$f_\rB(x) \approx 2{\sqrt{B^2-x^2}}$ \cite{ll}, and leads to a simpler procedure. 

Since weak coupling is large $B$, we study the resolvent in a systematic expansion in $1/B$. 
To do this, we consider the resolvent in two different regimes. In the {\it bulk regime} we take the limit
 \be
 \label{Btheta}
 B, \, \, x  \rightarrow \infty, \qquad u={x \over B} \, \, \, \text{fixed}.
 \ee
In the {\it edge regime} we consider instead 
 \be
B, \, \, x \rightarrow \infty,  \qquad  z=2\left(x-B\right) \, \, \, \text{fixed}.
 \ee

 For the bulk regime, and inspired by \cite{hutson,popov,iw-gy,volin}, we propose the following ansatz in the LL case, 
\be
\label{sol-ll}
\ba
& \tilde R_\rB (x)= -2 \pi {\sqrt{x^2-B^2}} \\
&+\sum_{n=0}^\infty \sum_{m=0}^\infty \sum_{k=0}^{n+m+1} {c_{n,m,k}(x/B)^{p_\rB(k)} \over B^{m-n-1}\left(x^2-B^2\right)^{n+{1\over 2}}}\ell^k(x,B), 
\ea
 \ee
 where $p_\rB(k)=k$ mod $2$, and %
 \be
 \ell(x, B)=\log\left( {x\!-\!B \over x\!+\!B} \right). 
 \ee
 In the GY case, it is given by 
\be
\label{sol-gy}
\ba
& R_\rF (x)= -\log\left( {x-\!B \over x+\!B} \right) \\
&+\sum_{n=1}^\infty \sum_{m=0}^\infty \sum_{k=0}^{n+m} {c_{n,m,k}(x/B)^{p_\rF (k)} \over B^{m-n}\left(x^2-B^2\right)^{n}}\ell^k(x,B),
\ea 
 \ee
where $p_\rF(k)=k-1$ mod $2$. Note that they are both odd functions on $\IR\backslash [-B, B]$. 
By computing their discontinuity we find an expansion for the density of roots similar to the one considered in \cite{hutson,popov,iw-gy}. 
The coefficients $c_{n,m,k}$ have all the information needed to compute the integrals (\ref{b-gam}), (\ref{b-e}), (\ref{f-gam}), (\ref{f-e}) as power series 
in $B$. For example, in the LL model, one has 
\be
\label{capac}
 \int_{-B}^B f_B(x) \rd x = \pi B^2+B \sum_{m=0}^\infty \frac{c_{0,m,0}-2 c_{0,m,1}}{B^{m}}. 
 \ee
 The $c_{n,m,k}$ can be partially determined by using the difference equations (\ref{gy-ll}), but this is not enough to fix their value. We then look at the edge regime. 

To study the edge regime, we write $R(z)$ as a Laplace transform, 
\be
R(z) = \int_0^\infty \hat R(s) \re^{-s z} \rd s. 
\ee
The difference equations (\ref{gy-ll}) become discontinuity equations along the negative $s$-axis:
\be
\label{laplaceequation}
\ba
 \sin(s)\left(\re^{-\ri s}\hat{R}_\rB (s^-)+\re^{\ri s}\hat{R}_\rB (s^+)\right)&=0, \\
\cos(s) \left( \re^{-\ri s} \hat R_\rF (s^-)-\re^{\ri s} \hat R_\rF(s^+) \right)&={1\over s^{-}} -{1\over s^{+}}, 
\ea 
\ee
where $s^\pm =s \pm \ri \epsilon$. We have simply denoted by $\hat R_\rB (s)$ the inverse Laplace transform of $\tilde R_\rB (z)$. 
Order by order in a $1/B$ expansion, the functions $\hat R(s)$ have to be analytic everywhere except on the negative real axis, and they must have an expansion as $s\rightarrow \infty$ in integer, 
negative powers of $s$ (this can be justified as in \cite{volin, volin-thesis}). The structure of poles and zeroes can be read from (\ref{laplaceequation}). 
The most general solution of (\ref{laplaceequation}) which satisfies these properties is the following:
 \be
 \label{sol1}
  \hat R(s)=   \Phi(s) \left( {1\over s} + Q(s) \right),
  \ee
  where 
\be
\ba
\Phi_\rB(s)&={ {\sqrt{ \pi B}} \over {\sqrt{s}}} \exp \left[ {s \over  \pi}  \log\left({ \pi \re \over  s}\right)\right] \Gamma\left( {s\over  \pi}+1 \right),\\
\Phi_\rF(s)&= {1\over {\sqrt{\pi}}} \exp \left[ {s \over  \pi}  \log\left({ \pi \re \over  s}\right)\right] \Gamma\left( {s\over  \pi}+{1\over 2} \right),
\ea
\ee
and
  \be
  \ba
  Q_\rB (s)&={1\over B s} \sum_{m=0}^\infty \sum_{n=0}^{m+1}  { Q^\rB_{n,m+1-n}(\log B) \over B^{m} s^{n}}, \\ Q_\rF (s)&=
  {1\over Bs} \sum_{m=0}^\infty \sum_{n=0}^{m}  { Q^\rF_{n,m-n}(\log B) \over B^{m} s^{n}}. 
   \ea
\ee
The coefficients $Q^{\rB, \rF}_{n,m}$ are undetermined functions of $\log B$. 
Finally, we match the bulk and the edge regimes as in \cite{hutson, popov, volin}: we first consider the ansatz (\ref{sol-ll}) and (\ref{sol-gy}) for $R_{\rB, \rF}(x)$ in the bulk regime, 
we set $x=z/2 +B$, and we expand around $z =\infty$ and $B=\infty$. We do the Laplace transform of the resulting expression, and we compare the result to the expansion of (\ref{sol1}) around $s=0$. This turns out to fix all the coefficients $c_{n,m,k}$ and $Q_{n,m}$, therefore the ground state energy density of the LL and the GY models. 

An example will illustrate how this works. By expanding (\ref{sol-ll}), we find
 \be
 \tilde R_\rB (z)=-2\pi\sqrt{B z}+\frac{\sqrt{B}}{\sqrt{z}}\left(c_{0,0,0}+ c_{0,0,1}\log\left({z\over 4B}\right)\right)+\cdots
 \ee
 The expansion of $\Phi_\rB(s)/s$ at $s=0$ reads, 
\be
{\Phi_\rB(s) \over s}={\sqrt{\pi B}} \left( \frac{1}{s\sqrt{s}}+\frac{1-\gamma_E+\log\left(\frac{\pi}{s}\right)}{\pi \sqrt{s}}+\cdots \right), 
\ee
and the Laplace transform of (\ref{sol1}) gives
\be
\label{fg-edge}
 \tilde R_\rB (z)=  {\sqrt{\pi B}} \left( - 2\sqrt{\pi z} + \frac{1+\log(4 \pi z)}{\sqrt{\pi z}}+\cdots\right).
\ee
By comparing the coefficients we obtain,
\be
c_{0,0,0}=1+ \log(16 \pi B), \qquad c_{0,0,1}=1. 
\ee

The recursive procedure to determine the coefficients $c_{n,m,k}$ can be fully automatized, as in \cite{volin, volin-thesis}, and one can push the 
calculation to any desired order, the only limitation being CPU time. We obtain in this way two series in $1/B$, one for $\gamma$ and another for $\langle x^2\rangle$, whose coefficients are polynomials in $\log B$. 
When we re-express $e_{\rB, \rF}(\gamma)$ in terms of $\gamma$, logarithms disappear in the final result, and one obtains a power series in $\gamma^{1/2}$ for the LL case, and in $\gamma$ for the GY case. In the LL case, we find
\be
\ba
&e_\rB (\gamma)=\gamma -\frac{4 \gamma ^{3/2}}{3 \pi }+\left(\frac{1}{6}-\frac{1}{\pi ^2}\right) \gamma ^2
+\frac{3 \zeta (3)-4}{8 \pi ^3}\gamma
   ^{5/2} \\ 
   &+\frac{3 \zeta (3)-4}{24 \pi ^4}\gamma ^3+\frac{60
   \zeta (3)-45 \zeta (5)-32}{1024 \pi ^5}\gamma ^{7/2}\\
   &+\frac{3 \left(4 \zeta (3)+6 \zeta (3)^2-15 \zeta
   (5)\right)}{2048 \pi ^6}\gamma ^4 \\
   &+\frac{-4368 \zeta (3)+6048 \zeta (3)^2+2520 \zeta (5)-8505 \zeta
   (7)+1024 }{786432 \pi ^7}\gamma ^{9/2} \\
   &+\CO(\gamma^{5}), 
   \ea
   \ee
  while in the GY case we find, 
\be
\label{eten}
\ba
&e_\rF (\gamma)= \frac{\pi ^2}{12}-\frac{\gamma}{2}-\frac{\gamma^2}{12}-\frac{ \zeta (3)}{\pi ^4}\gamma^3-\frac{3 
   \zeta (3)}{2 \pi ^6}\gamma^4-\frac{3  \zeta (3)}{\pi ^8}\gamma^5\\
   &-\frac{5  (5 \zeta (3)+3 \zeta
   (5))}{4 \pi ^{10}}\gamma^6-\frac{3  \left(35 \zeta (3)+12 \zeta (3)^2+75 \zeta
   (5)\right)}{8 \pi ^{12}}\gamma^7\\
   &-\frac{63  \left(7 \zeta (3)+12 \zeta (3)^2+35 \zeta (5)+12
   \zeta (7)\right)}{16 \pi ^{14}}\gamma^8\\
   &-\frac{3 \left(404 \zeta (3)^2+\zeta (3) (240 \zeta (5)+77)+147 (5 \zeta (5)+6 \zeta (7))\right)}{4 \pi
   ^{16}} \gamma^9\\
   &+ \CO\left(\gamma^{10}\right).
   \ea
   \ee
We have calculated the exact form of the first $50$ coefficients for both the LL and the GY series. Our results are in accord with the numerical calculations and conjectures 
of \cite{prolhac,lang,zr-numerical} for the very first coefficients. As a bonus, we note that (\ref{capac}) is (up to an overall factor) the 
asymptotic expansion of the capacity of the circular plate condenser, a problem which goes back 
to the XIXth century \cite{hutson}. The method explained in this paper gives a systematic and explicit procedure to determine this expansion. 

\section{\label{lob}Large order behavior}

One of the motivations for this work is the study of the large order behavior of the perturbative series in quantum many-body systems. Since the pioneering work of 
Bender and Wu \cite{bw2}, we know that this behavior provides crucial information on the non-perturbative structure of the model. This 
connection between perturbative and non-perturbative sectors has now evolved 
into the theory of resurgence (see \cite{mmlargen,abs} for reviews). Although the large order behavior of perturbation theory has been 
investigated in Fermi systems in e.g. \cite{baker-review,rossi}, we know comparatively less than in quantum mechanics or even quantum field theory, due in part to the scarcity of 
examples where the perturbative series can be computed to high order. Our results on the LL and the GY models provide enough data to determine with some precision the large order 
behavior of the series for the ground state energy density. Let us denote,
\be
e_\rB(\gamma) = \gamma \sum_{k \ge 0} c^\rB_k \gamma^{k/2}, \qquad 
e_\rF(\gamma) = \sum_{k \ge 0} c^\rF_k \gamma^k.
\ee
One finds, for $k \gg 1$, 
\be
\label{large-order}
\ba
c^{\rB}_k &\sim  \left( 8 \pi \right)^{-k} \Gamma(k), \\ 
c^{\rF}_k & \sim (\pi^2)^{-k+1} \Gamma(k-1). 
\ea
\ee
\begin{figure}[ptb] 
\begin{center} 
\includegraphics[height=3.5cm] {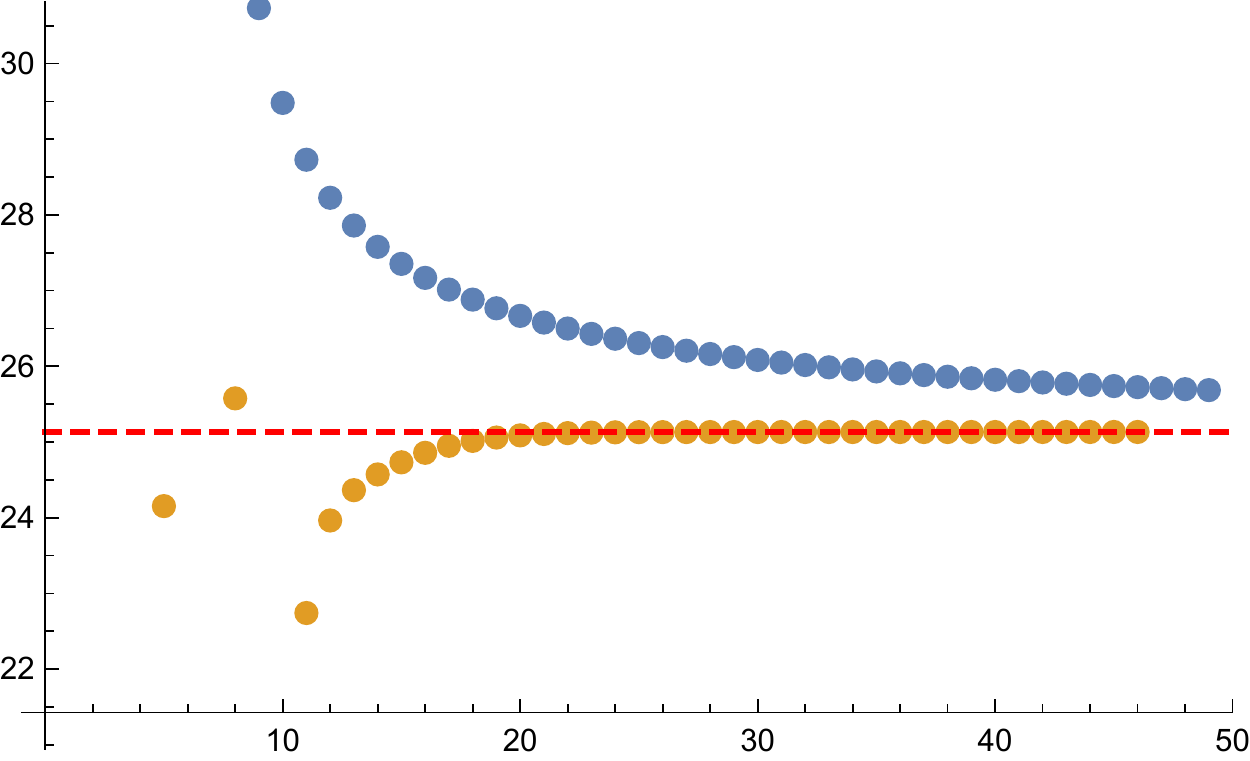}
\end{center} 
\caption{The upper dots represent the sequence $k c^{\rB}_k/c^{\rB}_{k+1}$, while the bottom dots represent its third Richardson transform. According to (\ref{large-order}), these 
sequences should converge to the value $8\pi$, represented here by a horizontal dashed line.}
\label{ll-as-fig} 
\end{figure}

The values $8 \pi$, $\pi^2$ have been guessed from the numerics with a precision of six decimal digits, see \figref{ll-as-fig} for an illustration in the LL case. We see that 
both series are factorially divergent and non-alternating, and there are singularities in the positive real axis of the Borel plane at $8 \pi$ (for LL) and $\pi^2$ (for GY). Therefore, the 
series are not Borel summable. However, one can perform lateral Borel resummations by deforming the integration contour slightly above 
or below the positive real axis. By standard arguments (see e.g. \cite{mmbook}), the resulting resummations have an ambiguous imaginary part which is 
exponentially small. It follows from (\ref{large-order}) that these non-perturbative ambiguities are of the form
\be
\label{ambi}
\text{LL}: \, \, \re^{-8 \pi/\gamma^{1/2}}, \qquad \text{GY}: \re^{-\pi^2/\gamma}. 
\ee
The perturbative series for the attractive GY model is obtained by changing $\gamma \rightarrow -\gamma$ (note however that the exact energies are not related by this transformation). 
The resulting series is therefore alternating and very likely it is Borel summable. In addition, its Borel resummation seems to agree with the exact result (see \cite{mr-long}). 

The result for the non-perturbative ambiguity in the GY model is particularly interesting. Instanton estimates indicate that, in fermionic theories, 
the growth of the perturbative series should be slower than factorial \cite{parisi-fermi,baker-pirner}, due to cancellations among diagrams. Our explicit 
results show that, at least in the GY model, this does not occur, and suggest that the divergence of the perturbative series can not be explained 
by semiclassical effects. In addition, one observes that the non-perturbative ambiguity in the GY model has the same 
exponential dependence than the square of the superconducting or spin gap of the model, which 
goes like $\Delta\sim \re^{-\pi^2/(2 \gamma)}$ \cite{ko,frz}. This is consistent with the fact that, in BCS theory, 
the ground state energy receives corrections of order $\Delta^2$. In \cite{mr-long} we argue that this behavior might be universal, namely, 
that the large order behavior of the perturbative series in weakly interacting Fermi systems is closely related to the superconducting gap. We also argue that 
this behavior is due to the factorial growth of certain classes of diagrams, akin to the renormalon singularity in quantum field theory.

\section{Conclusions}

In conclusion, in this paper we have solved the longstanding problem of determining the perturbative series of the ground state energy density in the LL and GY models, from 
the exact Bethe ansatz solution. The method we have developed, building on \cite{volin, volin-thesis}, is likely to have many other applications (see \cite{mr-long} for some illustrations). 
The results on the large order behavior of the perturbative series open the 
possibility to study their relation to non-perturbative phenomena, and more generally to apply the tools of the theory of resurgence in many-body physics. 
In the case of the GY model, we have argued that the non-perturbative ambiguity is closely related to the superconducting energy gap, providing in this way a new perspective on 
the Cooper instability. It would be very interesting to unravel the precise source of the non-perturbative ambiguity in the LL model, and to test recent resummation techniques like the one presented in \cite{rossi} in these
exactly solvable models. 

\vskip .2cm
{\it Acknowledgements:} We would like to thank 
Sylvain Prolhac, Wilhelm Zwerger and specially Thierry Giamarchi and F\'elix Werner for useful discussions and comments on the manuscript. This work is is supported in part by the 
Fonds National Suisse, subsidies 200021-175539, and by the NCCR 51NF40-182902 ``The Mathematics of Physics'' (SwissMAP).

\bibliographystyle{apsrev4-1} 
\bibliography{biblio-1d} 

\begin{thebibliography}{32}%
\makeatletter
\providecommand \@ifxundefined [1]{%
 \@ifx{#1\undefined}
}%
\providecommand \@ifnum [1]{%
 \ifnum #1\expandafter \@firstoftwo
 \else \expandafter \@secondoftwo
 \fi
}%
\providecommand \@ifx [1]{%
 \ifx #1\expandafter \@firstoftwo
 \else \expandafter \@secondoftwo
 \fi
}%
\providecommand \natexlab [1]{#1}%
\providecommand \enquote  [1]{``#1''}%
\providecommand \bibnamefont  [1]{#1}%
\providecommand \bibfnamefont [1]{#1}%
\providecommand \citenamefont [1]{#1}%
\providecommand \href@noop [0]{\@secondoftwo}%
\providecommand \href [0]{\begingroup \@sanitize@url \@href}%
\providecommand \@href[1]{\@@startlink{#1}\@@href}%
\providecommand \@@href[1]{\endgroup#1\@@endlink}%
\providecommand \@sanitize@url [0]{\catcode `\\12\catcode `\$12\catcode
  `\&12\catcode `\#12\catcode `\^12\catcode `\_12\catcode `\%12\relax}%
\providecommand \@@startlink[1]{}%
\providecommand \@@endlink[0]{}%
\providecommand \url  [0]{\begingroup\@sanitize@url \@url }%
\providecommand \@url [1]{\endgroup\@href {#1}{\urlprefix }}%
\providecommand \urlprefix  [0]{URL }%
\providecommand \Eprint [0]{\href }%
\providecommand \doibase [0]{http://dx.doi.org/}%
\providecommand \selectlanguage [0]{\@gobble}%
\providecommand \bibinfo  [0]{\@secondoftwo}%
\providecommand \bibfield  [0]{\@secondoftwo}%
\providecommand \translation [1]{[#1]}%
\providecommand \BibitemOpen [0]{}%
\providecommand \bibitemStop [0]{}%
\providecommand \bibitemNoStop [0]{.\EOS\space}%
\providecommand \EOS [0]{\spacefactor3000\relax}%
\providecommand \BibitemShut  [1]{\csname bibitem#1\endcsname}%
\let\auto@bib@innerbib\@empty
\bibitem [{\citenamefont {Lieb}\ and\ \citenamefont {Liniger}(1963)}]{ll}%
  \BibitemOpen
  \bibfield  {author} {\bibinfo {author} {\bibfnamefont {E.~H.}\ \bibnamefont
  {Lieb}}\ and\ \bibinfo {author} {\bibfnamefont {W.}~\bibnamefont {Liniger}},\
  }\href {\doibase 10.1103/PhysRev.130.1605} {\bibfield  {journal} {\bibinfo
  {journal} {Phys. Rev.}\ }\textbf {\bibinfo {volume} {130}},\ \bibinfo {pages}
  {1605} (\bibinfo {year} {1963})}\BibitemShut {NoStop}%
\bibitem [{\citenamefont {Gaudin}(1967)}]{gaudin}%
  \BibitemOpen
  \bibfield  {author} {\bibinfo {author} {\bibfnamefont {M.}~\bibnamefont
  {Gaudin}},\ }\href {\doibase https://doi.org/10.1016/0375-9601(67)90193-4}
  {\bibfield  {journal} {\bibinfo  {journal} {Phys. Lett.}\ }\textbf {\bibinfo
  {volume} {A24}},\ \bibinfo {pages} {55 } (\bibinfo {year}
  {1967})}\BibitemShut {NoStop}%
\bibitem [{\citenamefont {Yang}(1967)}]{yang}%
  \BibitemOpen
  \bibfield  {author} {\bibinfo {author} {\bibfnamefont {C.-N.}\ \bibnamefont
  {Yang}},\ }\href {\doibase 10.1103/PhysRevLett.19.1312} {\bibfield  {journal}
  {\bibinfo  {journal} {Phys. Rev. Lett.}\ }\textbf {\bibinfo {volume} {19}},\
  \bibinfo {pages} {1312} (\bibinfo {year} {1967})}\BibitemShut {NoStop}%
\bibitem [{\citenamefont {Jiang}\ \emph {et~al.}(2015)\citenamefont {Jiang},
  \citenamefont {Chen},\ and\ \citenamefont {Guan}}]{guan-boson}%
  \BibitemOpen
  \bibfield  {author} {\bibinfo {author} {\bibfnamefont {Y.-Z.}\ \bibnamefont
  {Jiang}}, \bibinfo {author} {\bibfnamefont {Y.-Y.}\ \bibnamefont {Chen}}, \
  and\ \bibinfo {author} {\bibfnamefont {X.-W.}\ \bibnamefont {Guan}},\ }\href
  {\doibase 10.1088/1674-1056/24/5/050311} {\bibfield  {journal} {\bibinfo
  {journal} {Chin. Phys. B}\ }\textbf {\bibinfo {volume} {24}},\ \bibinfo
  {pages} {050311} (\bibinfo {year} {2015})}\BibitemShut {NoStop}%
\bibitem [{\citenamefont {Guan}\ \emph {et~al.}(2013)\citenamefont {Guan},
  \citenamefont {Batchelor},\ and\ \citenamefont {Lee}}]{guan-review}%
  \BibitemOpen
  \bibfield  {author} {\bibinfo {author} {\bibfnamefont {X.-W.}\ \bibnamefont
  {Guan}}, \bibinfo {author} {\bibfnamefont {M.~T.}\ \bibnamefont {Batchelor}},
  \ and\ \bibinfo {author} {\bibfnamefont {C.}~\bibnamefont {Lee}},\ }\href
  {\doibase 10.1103/RevModPhys.85.1633} {\bibfield  {journal} {\bibinfo
  {journal} {Rev. Mod. Phys.}\ }\textbf {\bibinfo {volume} {85}},\ \bibinfo
  {pages} {1633} (\bibinfo {year} {2013})}\BibitemShut {NoStop}%
\bibitem [{\citenamefont {Cazalilla}\ \emph {et~al.}(2011)\citenamefont
  {Cazalilla}, \citenamefont {Citro}, \citenamefont {Giamarchi}, \citenamefont
  {Orignac},\ and\ \citenamefont {Rigol}}]{bthierry}%
  \BibitemOpen
  \bibfield  {author} {\bibinfo {author} {\bibfnamefont {M.~A.}\ \bibnamefont
  {Cazalilla}}, \bibinfo {author} {\bibfnamefont {R.}~\bibnamefont {Citro}},
  \bibinfo {author} {\bibfnamefont {T.}~\bibnamefont {Giamarchi}}, \bibinfo
  {author} {\bibfnamefont {E.}~\bibnamefont {Orignac}}, \ and\ \bibinfo
  {author} {\bibfnamefont {M.}~\bibnamefont {Rigol}},\ }\href {\doibase
  10.1103/RevModPhys.83.1405} {\bibfield  {journal} {\bibinfo  {journal} {Rev.
  Mod. Phys.}\ }\textbf {\bibinfo {volume} {83}},\ \bibinfo {pages} {1405}
  (\bibinfo {year} {2011})}\BibitemShut {NoStop}%
\bibitem [{\citenamefont {Bloch}\ \emph {et~al.}(2008)\citenamefont {Bloch},
  \citenamefont {Dalibard},\ and\ \citenamefont {Zwerger}}]{bdz}%
  \BibitemOpen
  \bibfield  {author} {\bibinfo {author} {\bibfnamefont {I.}~\bibnamefont
  {Bloch}}, \bibinfo {author} {\bibfnamefont {J.}~\bibnamefont {Dalibard}}, \
  and\ \bibinfo {author} {\bibfnamefont {W.}~\bibnamefont {Zwerger}},\ }\href
  {\doibase 10.1103/RevModPhys.80.885} {\bibfield  {journal} {\bibinfo
  {journal} {Rev. Mod. Phys.}\ }\textbf {\bibinfo {volume} {80}},\ \bibinfo
  {pages} {885} (\bibinfo {year} {2008})}\BibitemShut {NoStop}%
\bibitem [{\citenamefont {Popov}(1977)}]{popov}%
  \BibitemOpen
  \bibfield  {author} {\bibinfo {author} {\bibfnamefont {V.~N.}\ \bibnamefont
  {Popov}},\ }\href {https://doi.org/10.1007/BF01036714} {\bibfield  {journal}
  {\bibinfo  {journal} {Theor. Math. Phys.}\ }\textbf {\bibinfo {volume}
  {30}},\ \bibinfo {pages} {222} (\bibinfo {year} {1977})}\BibitemShut
  {NoStop}%
\bibitem [{\citenamefont {Iida}\ and\ \citenamefont {Wadati}(2005)}]{iw-gy}%
  \BibitemOpen
  \bibfield  {author} {\bibinfo {author} {\bibfnamefont {T.}~\bibnamefont
  {Iida}}\ and\ \bibinfo {author} {\bibfnamefont {M.}~\bibnamefont {Wadati}},\
  }\href@noop {} {\bibfield  {journal} {\bibinfo  {journal} {J. Phys. Soc.
  Jpn.}\ }\textbf {\bibinfo {volume} {74}},\ \bibinfo {pages} {1724} (\bibinfo
  {year} {2005})}\BibitemShut {NoStop}%
\bibitem [{\citenamefont {Tracy}\ and\ \citenamefont
  {Widom}(2016{\natexlab{a}})}]{tw-ll}%
  \BibitemOpen
  \bibfield  {author} {\bibinfo {author} {\bibfnamefont {C.~A.}\ \bibnamefont
  {Tracy}}\ and\ \bibinfo {author} {\bibfnamefont {H.}~\bibnamefont {Widom}},\
  }\href {\doibase 10.1088/1751-8113/49/29/294001} {\bibfield  {journal}
  {\bibinfo  {journal} {J. Phys.}\ }\textbf {\bibinfo {volume} {A49}},\
  \bibinfo {pages} {294001} (\bibinfo {year} {2016}{\natexlab{a}})}\BibitemShut
  {NoStop}%
\bibitem [{\citenamefont {Tracy}\ and\ \citenamefont
  {Widom}(2016{\natexlab{b}})}]{tw1}%
  \BibitemOpen
  \bibfield  {author} {\bibinfo {author} {\bibfnamefont {C.~A.}\ \bibnamefont
  {Tracy}}\ and\ \bibinfo {author} {\bibfnamefont {H.}~\bibnamefont {Widom}},\
  }\href {\doibase 10.1063/1.4964252} {\bibfield  {journal} {\bibinfo
  {journal} {J. Math. Phys.}\ }\textbf {\bibinfo {volume} {57}},\ \bibinfo
  {pages} {103301} (\bibinfo {year} {2016}{\natexlab{b}})}\BibitemShut
  {NoStop}%
\bibitem [{\citenamefont {Tracy}\ and\ \citenamefont {Widom}(2018)}]{tw2}%
  \BibitemOpen
  \bibfield  {author} {\bibinfo {author} {\bibfnamefont {C.~A.}\ \bibnamefont
  {Tracy}}\ and\ \bibinfo {author} {\bibfnamefont {H.}~\bibnamefont {Widom}},\
  }in\ \href@noop {} {\emph {\bibinfo {booktitle} {Geometric Methods in
  Physics}}},\ \bibinfo {editor} {edited by\ \bibinfo {editor} {\bibfnamefont
  {P.}~\bibnamefont {Kielanowski}}, \bibinfo {editor} {\bibfnamefont
  {A.}~\bibnamefont {Odzijewicz}}, \ and\ \bibinfo {editor} {\bibfnamefont
  {E.}~\bibnamefont {Previato}}}\ (\bibinfo  {publisher} {Springer},\ \bibinfo
  {year} {2018})\ pp.\ \bibinfo {pages} {201--212}\BibitemShut {NoStop}%
\bibitem [{\citenamefont {Prolhac}(2017)}]{prolhac}%
  \BibitemOpen
  \bibfield  {author} {\bibinfo {author} {\bibfnamefont {S.}~\bibnamefont
  {Prolhac}},\ }\href {\doibase 10.1088/1751-8121/aa5e00} {\bibfield  {journal}
  {\bibinfo  {journal} {J. Phys.}\ }\textbf {\bibinfo {volume} {A50}},\
  \bibinfo {pages} {144001} (\bibinfo {year} {2017})}\BibitemShut {NoStop}%
\bibitem [{\citenamefont {Lang}(2018)}]{lang}%
  \BibitemOpen
  \bibfield  {author} {\bibinfo {author} {\bibfnamefont {G.}~\bibnamefont
  {Lang}},\ }\href@noop {} {\emph {\bibinfo {title} {Correlations in
  low-dimensional quantum gases}}}\ (\bibinfo  {publisher} {Springer--Verlag},\
  \bibinfo {year} {2018})\BibitemShut {NoStop}%
\bibitem [{\citenamefont {Ristivojevic}(2019)}]{zr-numerical}%
  \BibitemOpen
  \bibfield  {author} {\bibinfo {author} {\bibfnamefont {Z.}~\bibnamefont
  {Ristivojevic}},\ }\href {\doibase 10.1103/PhysRevB.100.081110} {\bibfield
  {journal} {\bibinfo  {journal} {Phys. Rev. B}\ }\textbf {\bibinfo {volume}
  {100}},\ \bibinfo {pages} {081110} (\bibinfo {year} {2019})}\BibitemShut
  {NoStop}%
\bibitem [{\citenamefont {Volin}(2010)}]{volin}%
  \BibitemOpen
  \bibfield  {author} {\bibinfo {author} {\bibfnamefont {D.}~\bibnamefont
  {Volin}},\ }\href {\doibase 10.1103/PhysRevD.81.105008} {\bibfield  {journal}
  {\bibinfo  {journal} {Phys. Rev.}\ }\textbf {\bibinfo {volume} {D81}},\
  \bibinfo {pages} {105008} (\bibinfo {year} {2010})},\ \Eprint
  {http://arxiv.org/abs/0904.2744} {arXiv:0904.2744 [hep-th]} \BibitemShut
  {NoStop}%
\bibitem [{\citenamefont {Volin}(2011)}]{volin-thesis}%
  \BibitemOpen
  \bibfield  {author} {\bibinfo {author} {\bibfnamefont {D.}~\bibnamefont
  {Volin}},\ }\href {\doibase 10.1088/1751-8113/44/12/124003} {\bibfield
  {journal} {\bibinfo  {journal} {J. Phys.}\ }\textbf {\bibinfo {volume}
  {A44}},\ \bibinfo {pages} {124003} (\bibinfo {year} {2011})},\ \Eprint
  {http://arxiv.org/abs/1003.4725} {arXiv:1003.4725 [hep-th]} \BibitemShut
  {NoStop}%
\bibitem [{\citenamefont {Samaj}\ and\ \citenamefont {Bajnok}(2013)}]{sb-book}%
  \BibitemOpen
  \bibfield  {author} {\bibinfo {author} {\bibfnamefont {L.}~\bibnamefont
  {Samaj}}\ and\ \bibinfo {author} {\bibfnamefont {Z.}~\bibnamefont {Bajnok}},\
  }\href@noop {} {\emph {\bibinfo {title} {{Introduction to the statistical
  physics of integrable many-body systems}}}}\ (\bibinfo  {publisher}
  {Cambridge University Press},\ \bibinfo {year} {2013})\BibitemShut {NoStop}%
\bibitem [{\citenamefont {Ristivojevic}(2014)}]{zr}%
  \BibitemOpen
  \bibfield  {author} {\bibinfo {author} {\bibfnamefont {Z.}~\bibnamefont
  {Ristivojevic}},\ }\href {\doibase 10.1103/PhysRevLett.113.015301} {\bibfield
   {journal} {\bibinfo  {journal} {Phys. Rev. Lett.}\ }\textbf {\bibinfo
  {volume} {113}},\ \bibinfo {pages} {015301} (\bibinfo {year}
  {2014})}\BibitemShut {NoStop}%
\bibitem [{\citenamefont {Hutson}(1963)}]{hutson}%
  \BibitemOpen
  \bibfield  {author} {\bibinfo {author} {\bibfnamefont {V.}~\bibnamefont
  {Hutson}},\ }\href {\doibase 10.1017/S0305004100002152} {\bibfield  {journal}
  {\bibinfo  {journal} {Math. Proc. Cambridge Philos. Soc.}\ }\textbf {\bibinfo
  {volume} {59}},\ \bibinfo {pages} {211} (\bibinfo {year} {1963})}\BibitemShut
  {NoStop}%
\bibitem [{\citenamefont {Kostov}\ \emph {et~al.}(2008)\citenamefont {Kostov},
  \citenamefont {Serban},\ and\ \citenamefont {Volin}}]{ksv}%
  \BibitemOpen
  \bibfield  {author} {\bibinfo {author} {\bibfnamefont {I.}~\bibnamefont
  {Kostov}}, \bibinfo {author} {\bibfnamefont {D.}~\bibnamefont {Serban}}, \
  and\ \bibinfo {author} {\bibfnamefont {D.}~\bibnamefont {Volin}},\ }\href
  {\doibase 10.1088/1126-6708/2008/08/101} {\bibfield  {journal} {\bibinfo
  {journal} {JHEP}\ }\textbf {\bibinfo {volume} {08}},\ \bibinfo {pages} {101}
  (\bibinfo {year} {2008})},\ \Eprint {http://arxiv.org/abs/0801.2542}
  {arXiv:0801.2542 [hep-th]} \BibitemShut {NoStop}%
\bibitem [{\citenamefont {Bender}\ and\ \citenamefont {Wu}(1973)}]{bw2}%
  \BibitemOpen
  \bibfield  {author} {\bibinfo {author} {\bibfnamefont {C.~M.}\ \bibnamefont
  {Bender}}\ and\ \bibinfo {author} {\bibfnamefont {T.~T.}\ \bibnamefont
  {Wu}},\ }\href {\doibase 10.1103/PhysRevD.7.1620} {\bibfield  {journal}
  {\bibinfo  {journal} {Phys. Rev.}\ }\textbf {\bibinfo {volume} {D7}},\
  \bibinfo {pages} {1620} (\bibinfo {year} {1973})}\BibitemShut {NoStop}%
\bibitem [{\citenamefont {Mari{\~n}o}(2014)}]{mmlargen}%
  \BibitemOpen
  \bibfield  {author} {\bibinfo {author} {\bibfnamefont {M.}~\bibnamefont
  {Mari{\~n}o}},\ }\href {\doibase 10.1002/prop.201400005} {\bibfield
  {journal} {\bibinfo  {journal} {Fortsch. Phys.}\ }\textbf {\bibinfo {volume}
  {62}},\ \bibinfo {pages} {455} (\bibinfo {year} {2014})},\ \Eprint
  {http://arxiv.org/abs/1206.6272} {arXiv:1206.6272 [hep-th]} \BibitemShut
  {NoStop}%
\bibitem [{\citenamefont {Aniceto}\ \emph {et~al.}()\citenamefont {Aniceto},
  \citenamefont {Basar},\ and\ \citenamefont {Schiappa}}]{abs}%
  \BibitemOpen
  \bibfield  {author} {\bibinfo {author} {\bibfnamefont {I.}~\bibnamefont
  {Aniceto}}, \bibinfo {author} {\bibfnamefont {G.}~\bibnamefont {Basar}}, \
  and\ \bibinfo {author} {\bibfnamefont {R.}~\bibnamefont {Schiappa}},\
  }\href@noop {} {\ }\Eprint {http://arxiv.org/abs/1802.10441}
  {arXiv:1802.10441 [hep-th]} \BibitemShut {NoStop}%
\bibitem [{\citenamefont {Baker}(1971)}]{baker-review}%
  \BibitemOpen
  \bibfield  {author} {\bibinfo {author} {\bibfnamefont {G.~A.}\ \bibnamefont
  {Baker}},\ }\href {\doibase 10.1103/RevModPhys.43.479} {\bibfield  {journal}
  {\bibinfo  {journal} {Rev. Mod. Phys.}\ }\textbf {\bibinfo {volume} {43}},\
  \bibinfo {pages} {479} (\bibinfo {year} {1971})}\BibitemShut {NoStop}%
\bibitem [{\citenamefont {Rossi}\ \emph {et~al.}(2018)\citenamefont {Rossi},
  \citenamefont {Ohgoe}, \citenamefont {Van~Houcke},\ and\ \citenamefont
  {Werner}}]{rossi}%
  \BibitemOpen
  \bibfield  {author} {\bibinfo {author} {\bibfnamefont {R.}~\bibnamefont
  {Rossi}}, \bibinfo {author} {\bibfnamefont {T.}~\bibnamefont {Ohgoe}},
  \bibinfo {author} {\bibfnamefont {K.}~\bibnamefont {Van~Houcke}}, \ and\
  \bibinfo {author} {\bibfnamefont {F.}~\bibnamefont {Werner}},\ }\href
  {\doibase 10.1103/PhysRevLett.121.130405} {\bibfield  {journal} {\bibinfo
  {journal} {Phys. Rev. Lett.}\ }\textbf {\bibinfo {volume} {121}},\ \bibinfo
  {pages} {130405} (\bibinfo {year} {2018})},\ \Eprint
  {http://arxiv.org/abs/1802.07717} {arXiv:1802.07717 [cond-mat.quant-gas]}
  \BibitemShut {NoStop}%
\bibitem [{\citenamefont {Mari{\~n}o}(2015)}]{mmbook}%
  \BibitemOpen
  \bibfield  {author} {\bibinfo {author} {\bibfnamefont {M.}~\bibnamefont
  {Mari{\~n}o}},\ }\href@noop {} {\emph {\bibinfo {title} {Instantons and large
  $N$. An introduction to non-perturbative methods in quantum field theory}}}\
  (\bibinfo  {publisher} {Cambridge University Press},\ \bibinfo {year}
  {2015})\BibitemShut {NoStop}%
\bibitem [{\citenamefont {Mari\~no}\ and\ \citenamefont {Reis}()}]{mr-long}%
  \BibitemOpen
  \bibfield  {author} {\bibinfo {author} {\bibfnamefont {M.}~\bibnamefont
  {Mari\~no}}\ and\ \bibinfo {author} {\bibfnamefont {T.}~\bibnamefont
  {Reis}},\ }\href@noop {} {\ }\Eprint {http://arxiv.org/abs/1905.09569}
  {arXiv:1905.09569 [hep-th]} \BibitemShut {NoStop}%
\bibitem [{\citenamefont {Parisi}(1977)}]{parisi-fermi}%
  \BibitemOpen
  \bibfield  {author} {\bibinfo {author} {\bibfnamefont {G.}~\bibnamefont
  {Parisi}},\ }\href {\doibase 10.1016/0370-2693(77)90020-X} {\bibfield
  {journal} {\bibinfo  {journal} {Phys. Lett.}\ }\textbf {\bibinfo {volume}
  {66B}},\ \bibinfo {pages} {382} (\bibinfo {year} {1977})}\BibitemShut
  {NoStop}%
\bibitem [{\citenamefont {Baker}\ and\ \citenamefont
  {Pirner}(1983)}]{baker-pirner}%
  \BibitemOpen
  \bibfield  {author} {\bibinfo {author} {\bibfnamefont {G.~A.}\ \bibnamefont
  {Baker}, \bibfnamefont {Jr.}}\ and\ \bibinfo {author} {\bibfnamefont {H.~J.}\
  \bibnamefont {Pirner}},\ }\href {\doibase 10.1016/0003-4916(83)90334-2}
  {\bibfield  {journal} {\bibinfo  {journal} {Annals Phys.}\ }\textbf {\bibinfo
  {volume} {148}},\ \bibinfo {pages} {168} (\bibinfo {year}
  {1983})}\BibitemShut {NoStop}%
\bibitem [{\citenamefont {Krivnov}\ and\ \citenamefont
  {Ovchinnikov}(1975)}]{ko}%
  \BibitemOpen
  \bibfield  {author} {\bibinfo {author} {\bibfnamefont {V.~Y.}\ \bibnamefont
  {Krivnov}}\ and\ \bibinfo {author} {\bibfnamefont {A.}~\bibnamefont
  {Ovchinnikov}},\ }\href@noop {} {\bibfield  {journal} {\bibinfo  {journal}
  {J. Exp. Theor. Phys.}\ }\textbf {\bibinfo {volume} {40}},\ \bibinfo {pages}
  {781} (\bibinfo {year} {1975})}\BibitemShut {NoStop}%
\bibitem [{\citenamefont {Fuchs}\ \emph {et~al.}(2004)\citenamefont {Fuchs},
  \citenamefont {Recati},\ and\ \citenamefont {Zwerger}}]{frz}%
  \BibitemOpen
  \bibfield  {author} {\bibinfo {author} {\bibfnamefont {J.~N.}\ \bibnamefont
  {Fuchs}}, \bibinfo {author} {\bibfnamefont {A.}~\bibnamefont {Recati}}, \
  and\ \bibinfo {author} {\bibfnamefont {W.}~\bibnamefont {Zwerger}},\ }\href
  {\doibase 10.1103/PhysRevLett.93.090408} {\bibfield  {journal} {\bibinfo
  {journal} {Phys. Rev. Lett.}\ }\textbf {\bibinfo {volume} {93}},\ \bibinfo
  {pages} {090408} (\bibinfo {year} {2004})}\BibitemShut {NoStop}%
\end{thebibliography}%

\end{document}